\documentclass[a4paper,12pt]{iopart} % 12pt preprint style

\usepackage{iopams}   % sophisticated mathematics
\usepackage{harvard}  % bib style harvard
\usepackage{graphicx}

\newcommand{\Htwo}{{\mathrm{H}_2}}
\newcommand{\vect}[1]{{\mathbf{#1}}}
\newcommand{\Mrms}{{M_{\mathrm{rms}}}}
\newcommand{\MJ}{{M_\mathrm{J}}}
\newcommand{\LJ}{{\lambda_\mathrm{J}}}

\begin{document}

\title[Turbulent Mixing in the Interstellar Medium]{Turbulent Mixing in the Interstellar Medium --- an application for Lagrangian Tracer Particles}
\author{C Federrath $^{1,2}$, S C O Glover $^3$, R S Klessen $^1$ and W Schmidt $^4$}
\address{$^1$ Institut f\"ur Theoretische Astrophysik, Albert-Ueberle-Str.~2, 69120 Heidelberg, Germany}
\address{$^2$ Max-Planck-Institute for Astronomy, K\"onigstuhl 17, 69117 Heidelberg, Germany}
\address{$^3$ Astrophysikalisches Institut Potsdam, An der Sternwarte 16, 14482 Potsdam, Germany}
\address{$^4$ Institut f\"ur Theoretische Physik und Astrophysik, Am Hubland, 97074 W\"urzburg, Germany}
\ead{chfeder@ita.uni-heidelberg.de}

\begin{abstract}
We use 3-dimensional numerical simulations of self-gravitating compressible turbulent gas in combination with Lagrangian tracer particles to investigate the mixing process of molecular hydrogen ($\Htwo$) in interstellar clouds. Tracer particles are used to represent shock-compressed dense gas, which is associated with $\Htwo$. We deposit tracer particles in regions of density contrast in excess of $10\rho_0$, where $\rho_0$ denotes the mean density. Following their trajectories and using probability distribution functions, we find an upper limit for the mixing timescale of $\Htwo$, which is of order $0.3\,\mathrm{Myr}$. This is significantly smaller than the lifetime of molecular clouds, which demonstrates the importance of the turbulent mixing of $\Htwo$ as a preliminary stage to star formation.
\end{abstract}

\submitto{\PS (Conference Proceedings for ''Turbulent Mixing and Beyond 2007'')}

\section{INTRODUCTION}
Turbulent mixing of chemical species in the interstellar medium (ISM) has profound consequences on the morphology, chemistry, cooling and shielding of molecular clouds (MCs). MCs are known to be the formation site of young stars. It is therefore a prerequisite for a successful theory of star formation to explain the physical and chemical environment in which stars are born. Measured velocity dispersions suggest that interstellar clouds exhibit internal supersonic compressible random motions, that are associated with turbulence \citeaffixed{Larson1981,HeyerBrunt2004}{e.g.,}. Kinetic energy power spectra have been measured \citeaffixed{PadoanEtAl2006}{e.g.,}, which appear to be slightly steeper than the \citeasnoun{Kolmogorov1941a} spectrum of incompressible turbulence, even if intermittency corrections are considered \citeaffixed{Kolmogorov1962,SheLeveque1994,Boldyrev2002}{e.g.,}. Excellent summaries of empirical and theoretical aspects of interstellar turbulence are presented in the review articles by \citeasnoun{ElmegreenScalo2004}, \citeasnoun{ScaloElmegreen2004} and \citeasnoun{MacLowKlessen2004}.

Compressions created by the formation of strong shocks facilitate the formation of molecular hydrogen, $\Htwo$ \citeaffixed{HollenbachEtAl1971}{e.g.,}. Observations of MCs typically detect carbon monoxide as a tracer for $\Htwo$, which is found to be ubiquitous throughout MCs and not merely in regions of high density. Turbulent transport of $\Htwo$ from high density to low density gas may be an explanation for its relative homogeneity.

A recent numerical investigation of interstellar turbulence by \citeasnoun{GloverMacLow2007b} confirms that turbulent compressions are a key mechanism for the rapid ($1-2\,\mathrm{Myr}$) formation of $\Htwo$. Their results further suggest that a significant fraction of $\Htwo$ is located in low density gas, in quantities greater than can be formed directly {\em in situ}. The authors propose that $\Htwo$ forms in shock-compressed sheets and filaments, and is subsequently transported to lower density regions. In order to test this hypothesis, we have performed simulations of interstellar turbulence that employ Lagrangian tracer particles to directly follow the trajectories of dense gas parcels. Tracer particles are deposited in high density regions, formed in a driven compressible turbulent medium. We then use statistical analysis to show that even under self-gravitating conditions, significant mixing of $\Htwo$ occurs on short timescales ($<0.3\,\mathrm{Myr}$).

In section \ref{sec:method_and_setup}, we describe our numerical schemes and simulation setup. Section \ref{sec:results_discussion} presents our results and discussion, and in section \ref{sec:conclusions}, our conclusions are summarized.

\section{NUMERICAL METHODS AND SIMULATION SETUP} \label{sec:method_and_setup}
We solve the equations of compressible hydrodynamics including self-gravity on a 3-dimensional static grid of $256^3$ zones using the piecewise parabolic method (PPM) \cite{ColellaWoodward1984}, implemented in the astrophysical code ENZO \cite{OSheaEtAl2004} with periodic boundary conditions. Density $\rho$, velocity $\vect{v}$, total energy density $\rho e$, pressure $P$ and gravitational potential $\Phi$ are related through the equations
\begin{eqnarray}
\frac{\partial \rho}{\partial t} + \nabla \cdot(\rho \vect{v}) = 0 \label{eq:hydro1} \\
\frac{\partial \vect{v}}{\partial t} + (\vect{v} \cdot \nabla) \vect{v} = -\frac{1}{\rho}\nabla P - \nabla\Phi + \vect{f} \label{eq:hydro2} \\
\frac{\partial}{\partial t}(\rho e) + \nabla \cdot \bigl[\vect{v}(\rho e + P)\bigr] = -\rho\vect{v} \cdot (\nabla\Phi) + \rho \vect{v} \cdot \vect{f} \label{eq:hydro3} \\
\Delta\Phi = 4\pi G \rho \quad, \label{eq:poisson}
\end{eqnarray}
where $G$ is the gravitational constant. An isothermal equation of state,
\begin{equation}
P = \rho (\gamma - 1) u \qquad \mathrm{with} \qquad u = e - \frac{1}{2} \vect{v}^2 \label{eq:eos}\;,
\end{equation}
approximated by $\gamma=1.01$ is used to close the equations of hydrodynamics. This is a crude, but reasonable approximation on a wide range of scales in both length and density in interstellar clouds.

We start from a uniform distribution of gas, initially at rest. In order to excite turbulent motions, we utilize a stochastic forcing term $\vect{f}$, appearing as source term in equations (\ref{eq:hydro2}) and (\ref{eq:hydro3}) that supplies kinetic energy on the largest scales. The forcing term has a parabolic Fourier spectrum with $k/k_0 = [0.5,1.5]$ centered on $k_0 = 2\pi/L$, where $L$ is half of the size of the computational domain, and is evolved by an Ornstein-Uhlenbeck process \citeaffixed{EswaranPope1988,SchmidtEtAl2006}{e.g.,}. In this study, we adjust our forcing amplitude such that the fluid reaches a representative rms Mach number of $3$. We emphasize that our forcing is constructed such that we can regulate the relative strength of compressive modes ($\nabla \times \vect{f} = 0$) with respect to solenoidal modes ($\nabla \cdot \vect{f} = 0$). The ratio of kinetic energies at the injection scale,
\begin{equation} \label{eq:chi}
\chi = \frac{E_{\mathrm{sol}}}{E_{\mathrm{sol}} + E_{\mathrm{comp}}}
\end{equation}
is approximately $0.8$, which represents mostly solenoidal forcing. Note that $\chi$ cannot be exactly $1$, despite the utilization of purely solenoidal forcing, because supersonic flows always contain compressive modes. The consequences of varying the forcing amplitude and mode are discussed in \citeasnoun{SchmidtFederrath2007}.

Timescales are measured in units of the autocorrelation timescale $T$ of the forcing, which was set equal to the turbulent crossing time on large scales for a Mach $3$ turbulent medium. This means that after approximately $1\,T$, the gas reaches an rms Mach number $\Mrms \approx 3$, which is maintained by the forcing for all times $t \geq 1\,T$. Within $0 \leq t < 1\,T$, self-gravity is kept deactivated, allowing the gas to reach a state of fully developed compressible turbulence. After that, at $t=1\,T$, Lagrangian tracer particles are placed at the centers of grid cells with density $\rho/\rho_0 \geq 10$ ($\rho_0$ is the mean density), representing highly compressed gas parcels. From this point on, we distinguish three routes of further evolving the simulation:
\begin{itemize}
\item[1) ]{no changes, self-gravity still deactivated (pure forcing)}
\item[2) ]{self-gravity activated, representing 64 Jeans masses (forcing + SG4)}
\item[3) ]{self-gravity activated, representing 4096 Jeans masses (forcing + SG64) .}
\end{itemize}
Case 1 (pure forcing) serves as a control run to compare to the self-gravitating cases 2 and 3. The difference between 2 (SG4) and 3 (SG64) lies in the mean density $\rho_0$ of the fluid, which makes up a domain that contains a total of 64 Jeans masses for case 2 (SG4) and a total of 4096 Jeans masses for case 3 (SG64). The Jeans mass is defined by
\begin{equation}
\MJ = \rho_0 \LJ^3 = \rho_0 \Biggl( \frac{\pi c_s^2}{G \rho_0} \Biggr)^{3/2} \quad ,
\end{equation}
where $c_s = \sqrt{\gamma P / \rho}$ is the sound speed. This means that the self-gravitating cases represent gravitationally unstable gas, and both cases would be subject to gravitational collapse in the absence of turbulent fluctuations and stochastic forcing.

The Langrangian tracer particles are evolved every timestep (using a simple Euler-step) according to the Eulerian velocity, which is interpolated from the underlying grid by means of a second order accurate (triangular shaped cloud) method at the current position of each tracer particle. We have also tried a first order (cloud in cell) and a third order \citeaffixed{LekienMarsden2005}{tricubic,} spatial interpolation, and a predictor-corrector-step for time integration, which yielded no statistically significant differences.

\section{RESULTS AND DISCUSSION} \label{sec:results_discussion}

\begin{figure}[t]
\begin{center}
\includegraphics[width=0.5\linewidth]{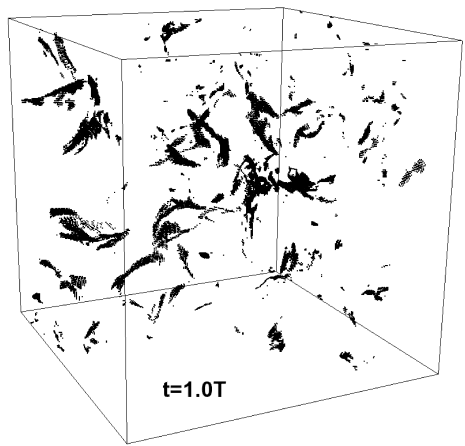}
\caption{Spatial distribution of tracer particles at the instance of their deposition $t=1.0\,T$ in regions with $\rho/\rho_0\geq10$. They represent shock-compressed gas, which is associated with $\Htwo$.} \label{fig:tracer_distrib_start}
\end{center}
\end{figure}

\begin{figure}[t]
\begin{center}
\includegraphics[width=0.8\linewidth]{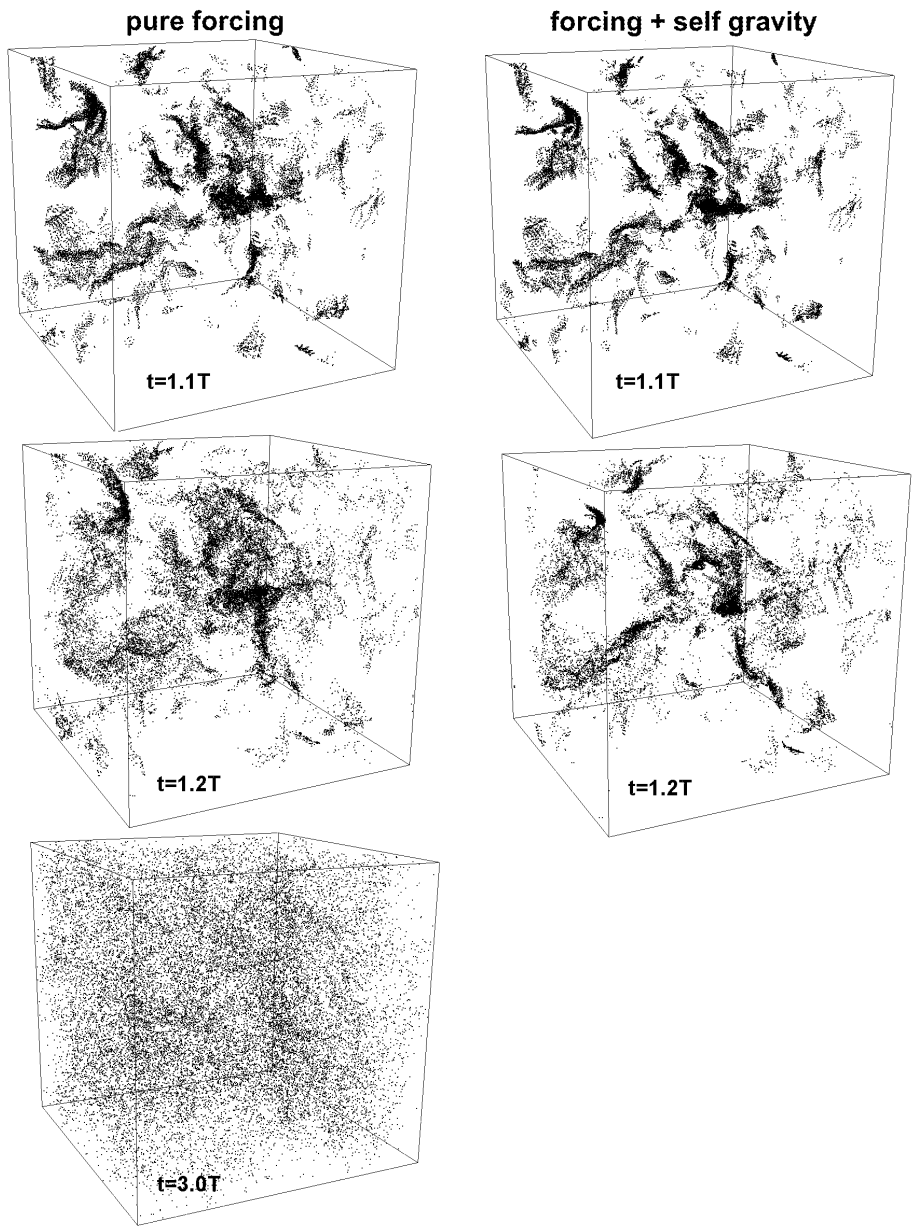}
\caption{Temporal evolution of the spatial distribution of Lagrangian tracer particles (from top to bottom). \textit{Left}: simulation without self-gravity (pure forcing). \textit{Right}: simulation including self-gravity (forcing + SG64). Self-gravity can prevent some of the tracer particles from mixing with the background gas. There are no snapshots for $t>1.2\,T$ for case SG64, because SG64 violates the resolution criterion for simulations including self-gravity \cite{TrueloveEtAl1997} shortly after $t=1.2\,T$, which only allows us to draw reliable conclusions up to this time. Case SG4 (not shown here) is statistically similar to the case of pure forcing, as discussed in the text.}
\label{fig:tracer_distrib_evol}
\end{center}
\end{figure}

Figure \ref{fig:tracer_distrib_start} shows the spatial distribution of tracer particles at $t=1.0\,T$, the instant of their deposition, which serves as an initial condition for three different routes of further development (see section \ref{sec:method_and_setup}). Approximately 24000 tracer particles have been placed in regions of shock compressed gas with density contrast in excess of $10$, representing $\Htwo$ in the following. In figure \ref{fig:tracer_distrib_evol}, we compare the evolution of the pure forcing run (left panel) to the 'strong' self-gravitating case SG64 (right panel). The case of 'weak' self-gravity (SG4) is not shown here, because it is almost identical to the evolution of the pure forcing run. This is because the forcing still dominates the dynamics in this case, whereas self-gravity becomes the dominant force in case SG64. Visual inspection of SG64 in comparison to pure forcing reveals that mixing of tracer particles occurs for both cases, but mixing is less efficient in the self-gravitating case. This is to be expected, since self-gravity acts to confine dense cores, within which the turbulent kinetic energy is less than the potential energy. Tracer particles located in such regions are trapped by self-gravity.

\begin{figure}[t]
\begin{center}
\includegraphics[width=0.9\linewidth]{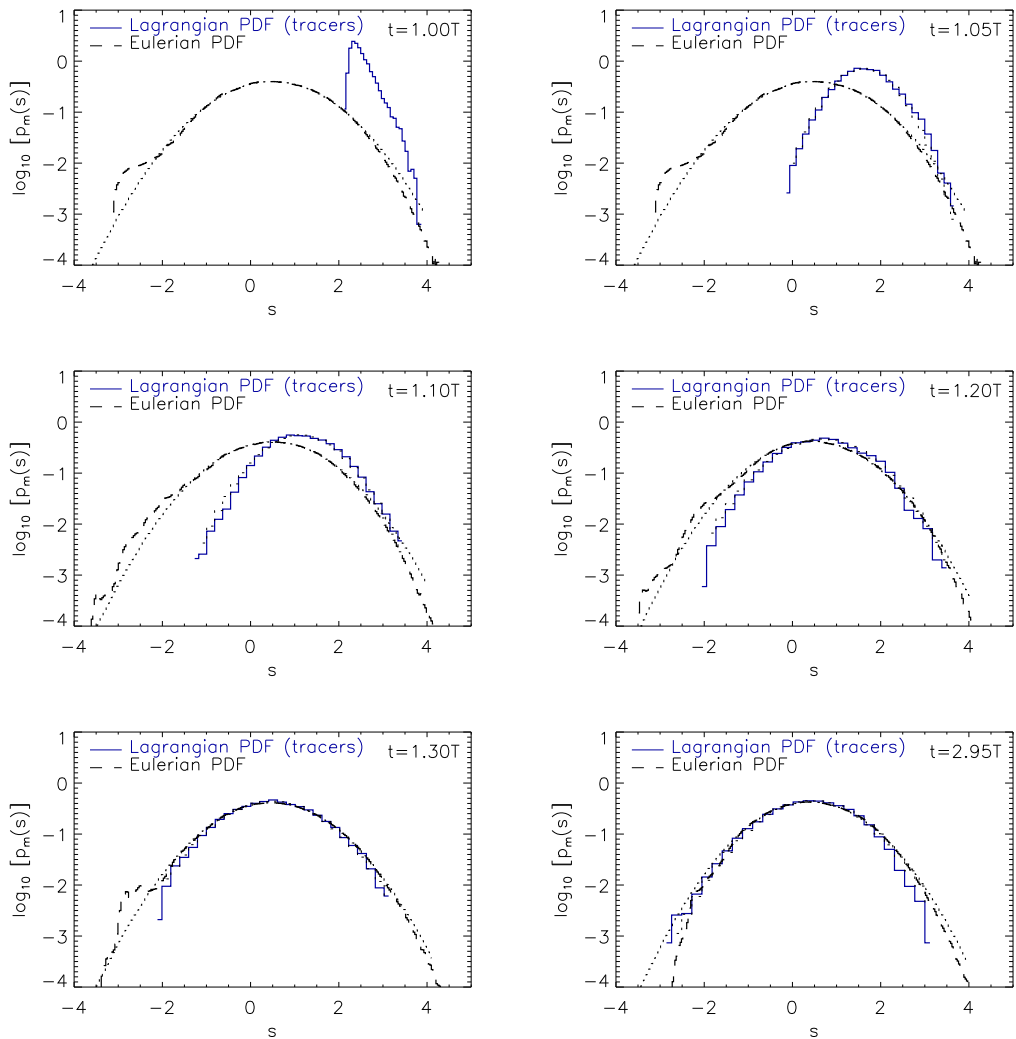}
\caption{Time sequence of the mass-weighted density PDFs for the case without self-gravity (pure forcing). The Lagrangian PDF of the tracer particles (solid line) gradually approaches the Eulerian PDF (dashed line). The dotted lines are least-squares-fits using the log-normal distribution given by equation (\ref{eq:log_normal_distrib}).} \label{fig:pdfs_nosg}
\end{center}
\end{figure}

\begin{figure}[t]
\begin{center}
\includegraphics[width=0.9\linewidth]{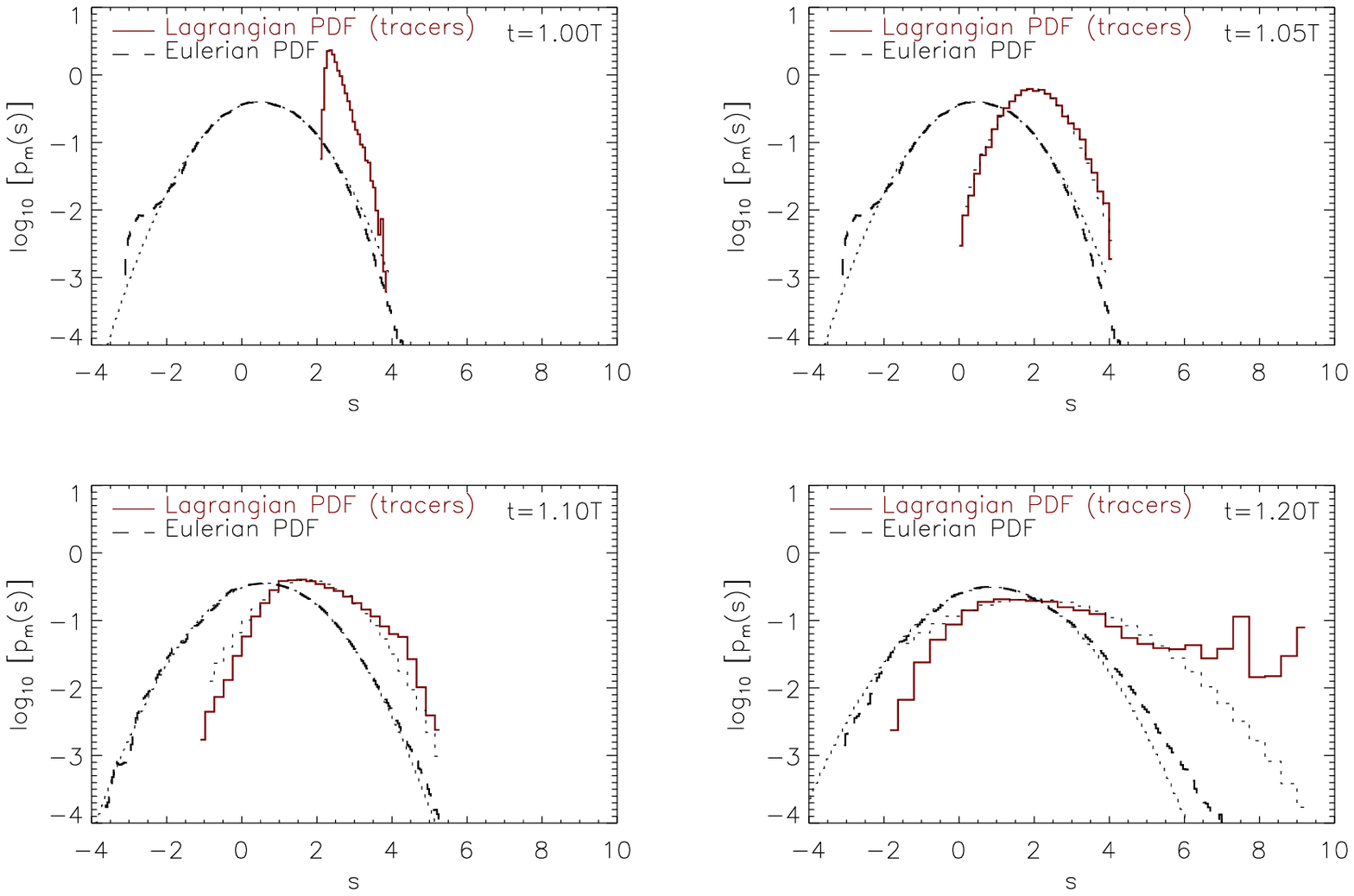}
\caption{Same as figure \ref{fig:pdfs_nosg}, but for the self-gravitating case SG64. The Lagrangian PDF of the tracer particles (solid line), as well as the Eulerian PDF (dashed line) deviate from a log-normal distribution at high densities and never fall on top of each other. In contrast to the case of pure forcing, complete mixing is impossible due to gravitational confinement. However, also in this case, turbulent transport to low density regions is very efficient. The dotted lines are least-squares-fits using the log-normal distribution given by equation (\ref{eq:log_normal_distrib}).} \label{fig:pdfs_sg64}
\end{center}
\end{figure}

In order to quantify this behavior, we compare probability distribution functions (PDFs) of the Eulerian gas density (density defined on the grid) and PDFs of Lagrangian gas density (interpolated from the grid and monitored at the position of each tracer particle). In fully developed compressible turbulent flows, the PDF of density is close to a log-normal distribution \citeaffixed{PadoanEtAl1997MNRAS,Klessen2000,OstrikerEtAl2001,LiEtAl2004,KritsukEtAl2007}{e.g.,} and given by
\begin{equation}
p_{v,m}(s)ds = \frac{1}{\sqrt{2\pi}\sigma} \mathrm{exp}\Biggl[-\frac{(s - s_{v,m})^2}{2 \sigma^2} \Biggr]ds \quad, \label{eq:log_normal_distrib}
\end{equation}
with $s \equiv \ln(\rho/\rho_0)$, mean $s_{v,m}$ and variance $\sigma^2$. The subscripts $v$ and $m$ denote volume- and mass-weighted distributions, respectively. Volume- and mass-weighted PDFs are related by $s_v = -s_m = -\sigma^2/2$ \citeaffixed{LiEtAl2003}{e.g.,}. We make use of this fact to transform the volume-weighted log-normal Eulerian PDF into the corresponding mass-weighted PDF. This allows us to directly compare the Eulerian PDF to the Lagrangian PDF of the tracer particles, which is naturally a mass-weighted distribution.

Figure \ref{fig:pdfs_nosg} shows a time sequence of the mass-weighted PDFs of both Eulerian (dashed lines) and Lagrangian (solid lines) gas density for the case of pure forcing. At $t=1.00\,T$, the tracer particle distribution covers densities in excess of $s \approx 2.3$, which corresponds to $\log_{10}[\rho/\rho_0] \approx 1$, as expected from their deposition criterion. The Eulerian distribution is in good agreement with a log-normal distribution, indicated by a fit using equation (\ref{eq:log_normal_distrib}) (dotted line), which justifies our transformation from volume- to mass-weighted PDFs. Following the time sequence, it is evident that the tracer particle PDF converges to the Eulerian PDF. For $t>1.30\,T$, both Eulerian and Lagrangian distributions are (except for small fluctuations in the wings of the distributions) statistically identical. This behavior indicates mixing of tracer particles to rarefied lower density regions.

In contrast to the case of pure forcing, the PDFs from SG64 shown in figure \ref{fig:pdfs_sg64} never reach a state where they resemble each other. Especially the high density wings, which are mostly affected by self-gravity deviate from a log-normal distribution. The Eulerian PDF tends to develop a power law tail for high density, in agreement with \citeasnoun{Klessen2000}. Although significant mixing into low density regions occurs within $0.1\,T$ of the tracer particle deposition, some tracers never completely mix with the background gas, owing to the self-gravity of the gas that holds them within deep gravitational potential wells in high density regions. The onset of gravitational collapse manifests in the increasing probability for tracer particles to be found at very high density, which is indicated by the flattening of the Lagrangian PDF towards high density.

\begin{figure}[t]
\begin{center}
\includegraphics[width=0.6\linewidth]{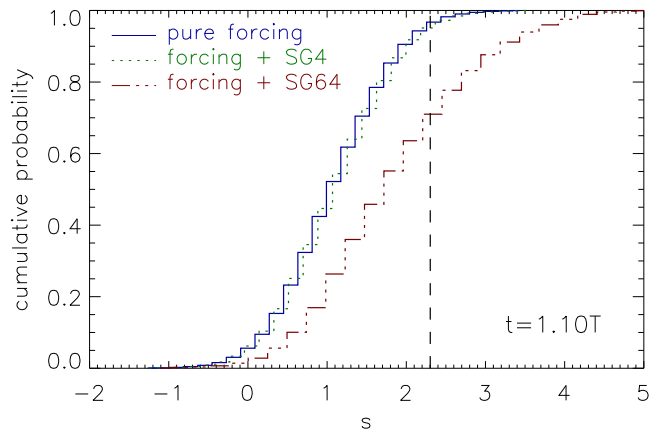}
\caption{Cumulative probability distribution of the density, which is monitored by Lagrangian tracer particles at $t=1.1\,T$ for the non-self-gravitating case (pure forcing, solid line), the 'weak' self-gravitating case (SG4, dotted line) and the 'strong' self-gravitating case (SG64, dash-dotted line). The vertical dashed line marks the density $\rho=10\rho_0$, which has been the deposition threshold for tracer particles at $t=1.0\,T$. All tracer particles, which contribute to the left of this line have been mixed to lower density regions. Pure forcing and SG4 are almost identical, because turbulence still dominates the dynamics. For SG64, $70\%$ of the tracer particles have been mixed to low density regions. For pure forcing and SG4, it is $95\%$.} \label{fig:cdfs}
\end{center}
\end{figure}

The fraction of tracer particles that have been mixed to regions of density smaller than their deposition density ($\rho < 10\rho_0$) can be estimated using cumulative distribution functions (CDFs) of the Lagrangian density distribution. We compare CDFs of all three runs (pure forcing, SG4 and SG64) at time $t=1.1\,T$ in figure \ref{fig:cdfs}. For the simulation without self-gravity and for SG4, approximately $95\%$ of the tracer particles have been mixed to regions of density smaller than $10\rho_0$ (vertical dashed line). In the case of 'strong' self-gravity (SG64), approximately $70\%$ have been transported to low density. This quantifies the influence of self-gravity in run SG64 on the mixing efficiency compared to pure forcing and SG4 cases. Although a significant fraction of tracer particles remains at high density, most of the tracer particles have already been mixed into lower density regions within $0.1\,T$ after their deposition, even in case SG64.

\section{CONCLUSIONS} \label{sec:conclusions}
Turbulent mixing of shock-compressed dense gas, which is associated with molecular hydrogen ($\Htwo$) and followed by means of Lagrangian tracer particles has been investigated in self-consistent 3-dimensional numerical simulations of driven supersonic turbulence. Even for self-gravitating media, turbulent transport from high density cores to low density rarefied regions is very efficient. Using Lagrangian tracer particles allowed us to directly confirm and quantify the importance of turbulent transport as the main mechanism for $\Htwo$ mixing in the interstellar medium \citeaffixed{GloverMacLow2007b}{see}. We estimate that approximately $70\%$ of the $\Htwo$ formed in shock-compressed clumps and filaments can be mixed into lower-density gas within one tenth of a turbulent crossing time for a solenoidally driven turbulent medium. This represents an upper limit for the mixing timescale of $0.3\,\mathrm{Myr}$ for typical interstellar cloud conditions in the cold neutral medium (rms Mach number $3$, cloud diameter $10\,\mathrm{parsec}$, sound speed $0.6\,\mathrm{km\,s^{-1}}$, dimensions similar to \citeasnoun{GloverMacLow2007b}). However, interstellar turbulence is likely to be driven compressively \citeaffixed[and references therein]{FederrathEtAl2007}{e.g.,}, a situation that is markedly different from solenoidally driven turbulence \cite{SchmidtFederrath2007}. This needs to be quantified in a follow-up study.

\ack
The simulations and post-processing used resources partly from a grant by the DEISA consortium and the HLRB II project grant h0972 at the Leibniz Supercomputing Centre Garching. C.~F.~partly received financial support from the SFB 439 'Galaxien im jungen Universum'. C.~F.~ is fellow of the Heidelberg Graduate School of Fundamental Physics and the International Max Planck Research School on Astronomy and Cosmic Physics at the University of Heidelberg.

\section*{References}
\bibliographystyle{jphysicsB}
\bibliography{turbulentmixing}

\end{document}